\documentstyle[preprint,prl,aps]{revtex}
\begin{document}
\preprint{}
\draft
\title{Exact Results for the \\ One-Dimensional Self-Organized Critical
Forest-Fire Model}
\author{Barbara Drossel, Siegfried Clar, and Franz Schwabl}
\address{Institut f\"ur Theoretische Physik, \\
         Physik-Department der Technischen Universit\"at M\"unchen, \\
         James-Franck-Str., D-85747 Garching, Germany}
\date{\today}
\maketitle
\begin{abstract}
We present the analytic solution of the self-organized critical (SOC)
forest-fire model in one dimension proving SOC in systems
without conservation laws by analytic means.
Under the condition that the system is in the steady
state and very close to the critical point, we calculate the
probability that a string of $n$ neighboring
sites is occupied by a given configuration of trees. The critical exponent
describing the size distribution of forest clusters is exactly
$\tau = 2$ and does not change under
certain changes of the model rules.
Computer simulations confirm the analytic results.
\end{abstract}


\narrowtext
The concept of self-organized criticality (SOC) \cite{bak88}
has attracted much attention during the last few years since it might
explain the origin of fractal structures and of $1/f$-noise in
nature. While the sandpile model \cite{bak88}, which is the
prototype for SOC, is well understood and has also been
investigated analytically \cite{dhar}, the investigation of
other SOC systems is mainly restricted to computer simulations
\cite{chr1,wol,tak1,bak2}. In particular the mechanism leading to SOC
in non conservative systems is barely understood. So far, there
exists no proof and even has been questioned
 that such non conservative systems can become critical.

A recent paper introduced  a critical forest-fire model which
is non conservative and showed that a double separation of
timescales leads to SOC
\cite{dro92}.  A series of computer simulations confirm the
criticality of the model \cite{gra92,chr93,hen93,cla93}, and a
scaling theory yields relations between various critical
exponents \cite{hen93,cla93}. An analytic proof for the
criticality of this model, however, has not been given so far.
\cite{chr93} contains the mean-field theory of the forest-fire
model, which seems to agree with simulations in high dimensions.
Critical exponents for the one-dimensional forest-fire model
have been derived by another mean-field approximation
in \cite{pac93}, but do not agree with simulations.

In this paper, we present the analytic solution of the
one-dimensional forest-fire model, thus proving for the first time that a
non conservative system can indeed show SOC. The analytic
calculation shows also that the value of the critical exponent
which characterizes the size distribution of forest clusters
does not change under certain changes of the model rules.
The values of the critical exponents have already been given in
\cite{dro92} by an argument which was rather intuitive and
failed in higher dimensions.

The forest-fire model \cite{dro92} is defined on a
$d$-dimensional hypercubic  lattice with
$L^d$ sites. Each site is occupied either by a tree or a burning
tree, or it is empty. The state of the system is parallely
updated by the following rules:
\begin{itemize}
\item[(i)] A burning tree becomes an empty site.
\item[(ii)] A green tree becomes a burning tree if at least one
of its nearest neighbors is burning.
\item[(iii)] A green tree becomes a burning tree with probability
$f\ll 1$  if no neighbor is burning.
\item[(iv)] At an empty site a tree grows with probability $p$.
\end{itemize}

A forest-fire model without rule (iii) was introduced
earlier \cite{bak90}.

After a transition period, the model assumes a steady state with
a constant mean forest density $\rho$.
Then the mean number of growing trees equals the
mean number of burning trees. During one timestep, there are on
an average $f\rho L^d$ lightning strokes in the system, and $p(1-
\rho)L^d$ trees grow. Consequently, the mean number of trees
destroyed by a lightning stroke is
\begin{equation}
\bar s=(f/p)^{-1}(1-\rho)/\rho.
\label{eq1}
\end{equation}
When the  tree growth parameter $p$ is small enough for a given
value $f/p$, forest clusters that are struck by lightning
burn down before new trees grow at their edge. In this case the
dynamics of the system are completely determined by the ratio $f/p$.
In the limit $f/p \to 0$, $\bar s$ diverges, and the system
approaches a critical point, where the size distribution of
forest clusters $n(s)$ obeys a
power law. Since any finite system is never exactly at the critical
point, the power law breaks down at a cluster size
$s_{\text{max}}\propto (f/p)^{-\lambda}$, and is given
by $n(s)\propto s^{-\tau}{\cal C}(s/s_{\text{max}})$.
The cutoff function ${\cal C}$ is ${\cal C}(x) = 1$ for $x < 1$
and ${\cal C}(x) \to 0$ for large $x$.
The critical behavior in the forest-fire model can
be observed independently of the initial conditions and over a
wide range of parameter values, as long as $f\ll p$ with sufficiently
small $p$, i.e. as long as time scales are separated. Then the
forest-fire model is {\em self-organized critical} \cite{bak88}.

We now turn to the one-dimensional case of this model and
investigate the properties of its steady state.
To avoid finite-size effects, we always assume that the system size
$L$ is $L \gg p / f$. The condition of time-scale separation reads
 in one dimension $p \ll f / p$. Consequently there are never two
lightning strokes or two growing trees at the same time within a
distance $\lesssim p / f$.
Let $n(s)$ be
the mean number of forest clusters of $s$ trees, divided by the
number of sites $L$. In \cite{dro92}, we already derived the
following relations
\begin{eqnarray}
\sum_{s=1}^\infty n(s) & = & (1-\rho-\rho f/p)/2, \nonumber \\
\sum_{s=1}^\infty sn(s) & = & \rho, \label{eq2} \\
\sum_{s=1}^\infty s^2n(s) & = & (1-\rho)p/f. \nonumber
\end{eqnarray}
The first relation follows from the condition that the total
number of forest clusters is constant in the steady state.

Before continuing, we need some definitions.
Each site can be in two possible states which we denote 0 (empty
site) and 1 (tree).
Let $i_\alpha \in \{0,1\}$ be the state of site  $\alpha$ and let
$P_n(i_1,\dots,i_n)$ be the probability that an arbitrarily
chosen string of $n$ sites is in the state $(i_1,\dots,i_n)$. We
also define the conditional probabilities $f_n(i_1,\dots, i_n)$ that
the configuration $(i_1,\dots, i_n)$ is ignited by fire during one
timestep.
$f_n(i_1,\dots, i_n)$ is the sum of the probability that a tree
belonging to the configuration $(i_1,\dots, i_n)$ is struck by
lightning, and, if $i_1 = 1$ or $i_n = 1$, the probability that
fire enters the configuration from outside.
In the following, we calculate the probabilities $P_n$ and $f_n$
neglecting terms which are smaller by a factor $f/p$  than the
leading terms. The results become more and more
exact when the critical point is approached.

The probabilities $P_1(i)$ can be expressed in terms of the mean
forest density: $P_1(1)=\rho$ and $ P_1(0)=1-\rho$.
At any given site, growth and burning of trees occur equally
often, and therefore
\begin{equation}
f_1(1)=p(1-\rho)/\rho.
\label{eq3}
\end{equation}
$f_1(1)$ is the sum of the probability that a tree is ignited
by a neighbor and the probability $f$ that it is struck by
lightning. The latter is smaller by a factor $f/p$ than
 $f_1(1)$ and therefore negligeable.

In addition to symmetry relations, the probabilities $P_2(i_1,i_2)$
satisfy $P_2(00)+P_2(10)=1-\rho $ and $P_2(10)+P_2(11)=\rho$.
The probabilities $f_2(i_1,i_2)$  are related by
$f_2(10)P_2(10)+f_2(11)P_2(11)=f_1(1)P_1(1)=p(1-\rho)$ and, since
the system is in the steady state, by
$f_2(11)P_2(11)=2pP_2(10)$ and $(p+f_2(10))P_2(10)=pP_2(00)$.
When a tree is ignited, the fire comes either from the right
neighbor or from the left neighbor.
This leads to $p(1-\rho)=f_2(11)P_2(11)$
and $f_2(10)=0$, and finally to \FL
\begin{equation}
P_2(00) = P_2(10) = (1 - \rho) / 2, \; P_2(11) = (3 \rho - 1) / 2.
\label{eq4}
\end{equation}
Remember that terms of order $f/p$ have been
neglected. In the limit $f/p\to 0$, i.e. at the critical point,
Eq.\ (\ref{eq4}) becomes exact. Before we consider the case $n\ge 3$,
we would like to comment the results Eq.\ (\ref{eq4}).
\begin{itemize}
\item[(i)] $P_2(10)=\sum_{s=1}^\infty n(s)$ (see (2)) since the
configuration 10 represents the right-hand edge of a forest cluster.
\item[(ii)] The exact result for $f_2(10)$ is
$f_2(10)P_2(10)=f\rho$ since each time  lightning strikes the system
a right-hand edge of a forest
cluster burns down.
$f_2(10)$ is smaller by a factor  $f/p$ than $f_2(11)$
and therefore negligeable since only
a portion $\propto f/p$ of all clusters burn down before trees
grow at their edge.
Below we will see that the approximation $f_2(10)=0$ leads to
$f_n(1\dots 1 0)=0$ for all $n>2$
simplifying considerably the calculations. $f_n(1\dots 1 0)=0$
cannot be true for large values of $n$ since forest clusters of
size $p/f$ are struck by lightning with a large probability,
which leads to $f_n(1\dots 1 0)\neq 0$. Our subsequent results
will therefore only be valid for $n \ll p/f$.
\end{itemize}

We are now able to calculate the probabilities $P_n$. From
$f_n(1 \dots 1) P_n(1 \dots 1) + f_n(1 \dots 10) P_n(1 \dots 10) =
f_{n - 1}(1 \dots 1) P_{n - 1}(1 \dots 1)$ and
$f_n(1 \dots 10) P_n(1 \dots 10) \le f_{n - 1}(1 \dots 10)
P_{n - 1}(1 \dots 10)$
we obtain by recursion
$f_n(1 \dots 1) P_n(1 \dots 1) = p(1 - \rho)$
and $f_n(1 \dots 10) = 0$. This means that a string of $n$ sites is
covered by a completely dense forest when it catches fire.
This dense forest belongs to a large forest cluster which has a
mean size $\bar s \gg n$ at the moment when it is struck by
lightning.
As long as not all sites are occupied by trees, the dynamics of
our string are completely determined by tree growth. All
configurations which contain the same number of trees therefore
have the same probability. Let $P_n(m)$ be the probability that
the string is occupied by $m$ trees. In the steady state the
$P_n(m)$ are related by the equations
\begin{eqnarray*}
p n P_n(0) & = & f_n(1\dots 1) P_n(n)=p(1-\rho), \\
p(n-m)P_n(m) & = & p(n-m+1)P_n(m-1)\;
\text{for}\; m \neq 0,n,
\end{eqnarray*}
which lead to the result
\begin{eqnarray}
P_n(m) & = & (1-\rho)/(n-m)\; \text{for}\;m<n, \nonumber \\
P_n(n) & = & 1-(1-\rho)\sum_{m=0}^{n-1}1/(n-m) \label{eq5} \\
& = & 1-(1-\rho) \sum_{m=1}^n 1/m. \nonumber
\end{eqnarray}
The size distribution of forest clusters is
\begin{eqnarray}
n(s) & = & P_{s+2}(01\dots 10)={P_{s+2}(s)\over
{s+2 \choose s}} \nonumber \\
& = & {1-\rho\over (s+1)(s+2)} \simeq (1-\rho)s^{-2}.
\label{eq6}
\end{eqnarray}
This is a power law with the critical exponent $\tau=2$.
Eq.\ (\ref{eq6}) leads to $\sum_{s=1}^\infty n(s)= (1-\rho)/2$, in
agreement with Eq.\ (\ref{eq2}). The size distribution of fires is
$\propto sn(s)\propto s^{-1}$.
The size distribution $n_e(s)$ of clusters of empty sites is \FL
\begin{equation}
n_e(s)=P_{s + 2}(10 \dots 01) = {P_{s + 2}(2) \over
{s + 2 \choose 2}} = {2 (1 - \rho) \over s (s + 1) (s + 2)},
\label{eq7}
\end{equation}
which is also derived in \cite{pac93}.

There is a characteristic length $s_{\text{max}}$
where the power law $n(s)\propto s^{-2}$ breaks down. During
one timestep, a cluster of size $s$ grows with
probability $2p$ to a bigger cluster, and it is destroyed by
lightning with probability $fs$. When the cluster size approaches
$s\propto p/f$, the cluster is struck by lightning with a finite
probability $fs$. This simple argument leads to
$s_{\text{max}}\propto p/f$ and $\lambda=1$, as did
the scaling theory in \cite{dro92}.
The remaining critical exponents are $\mu = \nu = 1$, as
already derived in \cite{dro92}.
The following microscopic derivation
shows that the relation for $s_{\text{max}}$ acquires
a logarithmic correction factor. We calculate
$s_{\text{max}}$ from the
condition that a string of size $n
\le s_{\text{max}}$ is not
struck by lightning until all trees are
grown. When a string of size $n$ is completely empty at time
$t=0$, it will be occupied by $n$ trees after
$T(n)=(1/p)\sum_{m=1}^n 1/m \simeq \ln(n)/p$ timesteps on an
average. The mean number of trees after $t$ timesteps is $m(t)=n[1-
\exp(-pt)]$. The probability that lightning
strikes a string of size $n$ before all trees are grown is
$f\sum_{t=1}^{T(n)} m(t)\simeq (f/p)n(\ln(n)-1) \simeq
(f/p)n\ln(n)$. We conclude
\begin{equation}
s_{\text{max}}\ln(s_{\text{max}})
\propto p/f \; \text{for large $p/f$}.
\label{eq8}
\end{equation}

Next we calculate the relation between the mean forest density
$\rho$ and the parameter $f/p$.
The probability $P_n(n)$ in Eq.\ (\ref{eq5}) cannot be less
than zero.  Since
$\sum_{m=1}^n 1/m\simeq \ln(n)$  for large values of $n$, the mean
forest density must approach the value one at the critical point
via $1-\rho \propto 1/\ln(s_{\text{max}}).$
A more precise result for $\rho$ is obtained from
\begin{eqnarray*}
\rho & = & \sum_{s=1}^\infty sn(s) \\
     & = & (1-\rho)\sum_{s=1}^{s_{\text{max}}}{s\over
           (s+1)(s+2)} + \sum_{s=s_{\text{max}}+1}^\infty sn(s) \\
     & \simeq & (1-\rho) \ln(s_{\text{max}}) +
                \int_{s_{\text{max}}}^\infty sn(s)ds.
\end{eqnarray*}
With the scaling ansatz $n(s) = (1-\rho)s^{-2} {\cal
C}(s/s_{\text{max}})$, the second term in
the last line reduces to $(1-\rho)$
multiplied by a constant factor, and thus
\begin{eqnarray}
{\rho \over 1-\rho} & = & \ln(s_{\text{max}})+
\text{const}. \simeq \ln(p/f)
 + \text{const}. \nonumber \\
& \simeq & \ln(p/f)\;
\text{for large $p/f$}.
\label{eq9}
\end{eqnarray}
The forest density approaches the value one at the
critical point. This is not
surprising since there exists no infinitely large
cluster in a one dimensional
system as long as the forest is not completely dense.
Combining Eqs.~(\ref{eq6}) and (\ref{eq9}), we obtain the
final result for the cluster-size distribution near the
critical point
\begin{equation}
n(s) \simeq \frac{1}{(s+1)(s+2)\ln s_{\text{max}}} \;
\text{for} \; s < s_{\text{max}}
\end{equation}
with $s_{\text{max}}$ given by Eq.~(\ref{eq8}).

Our computer simulations confirm the analytic results. The
simulations were performed using the same method as
in \cite{gra92}.
In Fig.~\ref{fig1} the size distribution $sn(s)$
of clusters struck by
lightning is plotted. The smooth
line is the analytic result derived from Eq.\ (\ref{eq6}).
It fits the simulation perfectly in the region
$s<s_{\text{max}}\simeq 1000$.
The bump in the cluster size distribution has the
following explanation: consider the dynamics of a forest
cluster. As long as the cluster size is $s < s_{\text{max}}$,
its average size increases each timestep by $2p(s+2)$
(can be derived from Eq.\ (\ref{eq5})), i.e. its growth
speed is $\dot{s} \propto s$. A tree spends a portion
$\propto 1/\dot{s}$ of its lifetime in clusters of size $s$,
and consequently the growth speed is related to the cluster
size distribution by
$sn(s) \propto 1 / \dot{s}$, which leads again
to the power law $n(s) \propto s^{-2}$. When the
cluster size becomes comparable to $s_{\text{max}}$, the
growth speed increases no longer proportionally to $s$
but slower than $s$ since the largest neighbors of the cluster
have already been destroyed by lightning.
Therefore $s^2n(s)$ increases with increasing $s$ for
$s > s_{\text{max}}$, giving rise to the bump in the
cluster size distribution. When the forest cluster
is  so large that it is struck by lightning
with a nonvanishing probability, $n(s)$
decreases exponentially.

The ratio $\ln(p/f)(1-\rho)/\rho$ is given in
Table~\ref{tab1} for different values of
$p/f$. It approaches the value 1 with increasing
$p/f$, as predicted by theory
(see Eq.\ (\ref{eq9}), the deviations from the value
1 for smaller $p/f$ are due to the
constant contribution in Eq.\ (\ref{eq9})). The ratio
$(p/f)/(s_{\text{max}}
\ln(s_{\text{max}}))$
is also shown in Table~\ref{tab1}
for different values of $p/f$. Within numerical
accuracy these ratios are
identical, in agreement with our theoretical result
Eq.\ (\ref{eq8}).

Finally, we calculate the temporal correlation function of the
number of burning sites. The mean number $N_s(t)$ of trees that
burn $t$
timesteps after a cluster of size $s$ is struck by lightning is
$N_s(t) \simeq 2(1-t/s)\theta(s-t)$.
The correlation function is then
\begin{eqnarray}
G(\tau) & \propto & \int_1^\infty \! ds \, n(s) s \int_0^{\infty}
\! dt \, N_s(t) N_s(t+\tau) \nonumber \\
& \propto & (2s_{\text{max}}/
\ln(s_{\text{max}})
 -3\tau(1-\ln(\tau)/\ln(s_{\text{max}}))). \label{eq10}
\end{eqnarray}
The Fourier transform of $G(\tau)$ is
$G(\omega)\propto \omega^{-2} (1+ \text{const.}
\ln(\omega s_{\text{max}}))$
 for small $\omega > 1 /
s_{\text{max}}$. There is a nontrivial deviation from the
$\omega^{-2}$ dependence in direction of $1/f$-noise.
 Our simulations confirm the power law dependence
 (Fig.~\ref{fig2}) but cannot discriminate between a
$\omega^{-2}$ law and the complete expression Eq.\ (\ref{eq10}).

We conclude with three comments.
\begin{itemize}
\item[(i)]
In the limit $f/p \to 0$ only an
infinitesimal portion of all forest clusters, all empty sites,
and even all trees are not described by our results
Eq.\ (\ref{eq5})-(\ref{eq7}) which therefore become exact at the
 critical point.
\item[(ii)] The self-organized critical forest-fire model is in
one dimension much simpler than in higher dimensions. The critical
exponents are classical in one dimension, and the tree
distribution on a string of size $n\le s_{\text{max}}$
is stochastic. In
higher dimensions, there are always trees left when a fire passes
through a region. Therefore the tree distribution is not
stochastic and the exponents are nontrivial.
\item[(iii)] The only conditions for obtaining the result
Eq.\ (\ref{eq5})
are that trees
grow stochastically and that a fire does not pass through a string
 of size $ n\le s_{\text{max}}$ as long as not all its trees are
grown. The $P_n$ and
consequently the cluster size distributions remain unchanged when
the rules for the occurrence of lightning are changed in one of
the following ways: (a) Lightning strikes all forest clusters of
size $\ge s_{\text{max}}$. This model is
critical in the limit $s_{\text{max}}\to \infty$. (b) Lightning
strikes every  $T$ timesteps the largest forest cluster on
each interval of $l$
sites. This version is critical for $l\to\infty$ with fixed $T$.
(c) Lightning strikes a
forest cluster of size $s$ with a probability $\epsilon f(s)$ which
 does not decrease with increasing  $s$. This model is critical for
$\epsilon\to 0$. While the size distribution of the forest clusters
and of the clusters of empty sites remains unchanged
in all these versions of the forest-fire model, the size
distribution of fires varies considerably and
is in general no power law.
\end{itemize}



\begin{table}
\caption{Numerical results for the ratios
${\ln(p/f) (1 - \rho)} / {\rho}$
and ${(p/f) / (s_{\text{max}}
\ln s_{\text{max}})}$ for different values of $p/f$.}
\begin{tabular}{cccccc}
$p/f$ & 3125 & 6250 & 12500 & 25000 & 50000 \\
\tableline
${\ln(p/f) (1 - \rho)} / {\rho}$ & 1.13 & 1.10 & 1.06 & 1.04 & 1.02 \\
${(p/f) / (s_{\text{max}}
\ln s_{\text{max}})}$ & 3.65 & 3.61 & 3.56 & 3.62 & 3.55 \\
\end{tabular}
\label{tab1}
\end{table}


\begin{figure}
\caption{Size distribution of the fires for $f/p=2/50000$ and $L=2^{20}$.
The smooth line is the theoretical result which is valid for cluster
sizes $\le s_{\text{max}}$.}
\label{fig1}
\end{figure}

\begin{figure}
\caption{The power spectrum $G({\omega})$ of the fire density. The
simulation parameters are $f/p=8/50000$ and $L=2^{20}$. The straight
line (theory) has the slope $-2$.}
\label{fig2}
\end{figure}


\end{document}